\documentclass[english,showpacs,twocolumn, superscriptaddress]{revtex4-1}
\usepackage[T1]{fontenc}
\usepackage[latin9]{inputenc}
\usepackage{textcomp}
\usepackage{amsmath}
\usepackage{amssymb}

\makeatletter
\usepackage[colorlinks = true,
            linkcolor =red,
            urlcolor  =blue,
            citecolor = blue,
            anchorcolor = blue]{hyperref}

\makeatother

\usepackage{babel}
\begin{document}
\title{Direct Measurement Methods of Density Matrix of an Entangled Quantum
State}
\author{Yusuf Turek}
\email{yusufu1984@hotmail.com}

\affiliation{School of Physics and Electronic Engineering, Xinjiang Normal University,
Urumqi, Xinjiang 830054, China}
\begin{abstract}
In general, the state of a quantum system represented by the density
operator and its determination is a fundamental problem in quantum
theory. In this study, two theoretical methods such as using postselected
measurement characterized by modular value and sequential measurements
of triple products of complementary observables to direct measurement
of matrix elements of density operator of a two photon entangled quantum
state are introduced. The similarity and feasibility of those two
methods are discussed by considering the previous experimental works. 
\end{abstract}
\maketitle

\section{Introduction}

In quantum mechanics the state can represent a quantum system, and
the improvement of the state and its determination has vital importance
in obtaining any information about that system. Because of the collapse
of wave function due to the decoherence, the conventional quantum
measurements cannot be directly used in some hot topics of quantum
information science such as quantum state based high precision measurements,
reconstruction of unknown quantum state, etc. However, the advance
of the research in fundamentals of quantum physics provided an effective
method to solve the above problems by using the simple and easily
manipulable pre- and post-selected quantum weak measurement technique
which is characterized by the weak value \citep{Aharonov1988}. In
the weak measurement the induced weak value of the observable on the
measured system is usually a complex number, and can be beyond the
usual range of eigenvalues of that observable. This property of weak
value is referred as the amplification effect for weak signal which
accompanied by the decrease of the postselection probability. Since
the weak signal amplification property experimentally demonstrated
in 1991 \citep{Ritchie1991}, it have been widely used and solved
plenty of fundamental problems in quantum mechanics and related sciences.
For details about the weak measurement theory and its applications
in weak signal amplification processes, we refer the reader to the
recent overview of the field \citep{KOFMAN201243,RevModPhys.86.307}.

Another main application of postselected weak measurement technique
is quantum state tomography. The significant advantageous of postselected
weak measurement based state tomography technique than conventional
one \citep{PhysRevA.40.2847,PhysRevLett.70.1244,PhysRevA.64.052312,RevModPhys.81.299,PhysRevLett.110.010404}
is that in weak measurement technique the tomographic procedures is
easy and can get the all global phase information of unknown state
than conventional schemes. Since J. Lundeen et al. \citep{Lundeen2011}
firstly investigated the reconstruction of transversal spatial wave
function of polarized photon beams by using the postselected weak
measurement technique, the direct measurement of unknown quantum states
have been studied theoretically and experimentally by using weak and
strong measurement techniques \citep{Lundeen2012,de_Gosson_2012,PhysRevA.86.052110,aBoyd2013,Sheng2013,Malik13,PhysRevLett.113.090402,Malik2014,PhysRevA.92.062133,PhysRevLett.115.090401,PhysRevA.91.052109,PhysRevLett.116.040502,PhysRevLett.117.170402,PhysRevA.93.052105,PhysRevA.93.062304,Thekkadath2016,PhysRevA.93.032128,PhysRevLett.118.010402,Calderaro2018}.
In particular, the direct measurement of a photon polarization state
in two dimensional system \citep{aBoyd2013} and direct measurement
of density matrix of a single photon polarization state in pure and
mixed state cases \citep{Thekkadath2016} showed the power of weak
measurement technique in state determination processes. 

Quantum entanglement is a main feature of quantum mechanics, and most
of the mysterious phenomena in quantum world caused by entangled systems.
Thus, the state determination of entangled systems have significant
importance in quantum theory. The direct measurement of general quantum
state by using weak measurement has been studied in Refs. \citep{Lundeen2012,Sheng2013}.
Furthermore, in recent innovative work of Guo-Guang Can et al. \citep{Pan2019},
they investigated the direct measurement of a two photon entangled
state by using postselected weak measurement and used the modular
value in reading results in stead of weak value. However, in general,
the state of a quantum system is represented by density operator,
and the direct measurement of density matrix of an entangled system
by using weak measurement technique have not been explicitly studied
until now.

In this paper, as an extension of previous works \citep{Thekkadath2016,Pan2019},
we study the two kinds of reconstruction methods of a two photon entangled
state. We take the spatial (paths) and polarization degrees of freedom
of unknown entangled state as pointer and measured system, respectively,
and the joint (or sequential ) projection operators of two subsystems
considered as measured observables of measured system. In first method,
we follow the theoretical part of Ref. \citep{Pan2019} and use the
postselected weak measurement technique to measure the matrix elements
of a two photon entangled state. It is noticed that the density matrix
elements proportional to the weak values of appropriate joint projection
operators of two subsystems, and the pre- and post-selected states
are the elements in two mutually unbiased bases. Since the weak measurement
of joint projection operators of two subsystems can not be measured
directly, it is founded in terms of the modular values of corresponding
operators. Based on the theoretical analysis of Ref. \citep{Pan2019},
the real and imaginary parts of a matrix element can be readout from
detection probability after taking appropriate projection operations
before detection on the final state of the pointer. 

In the second method, the technique introduced by J. Lundeen et al.
\citep{Lundeen2012} is used. Three sequential measurements on three
projection operators of two subsystems where each complementary to
the last are taken to find the matrix elements of a two photon entangled
state. It is found that the result of these sequential measurements
proportional to the value of matrix elements. In order to read out
the value of matrix elements, it is assumed that the spatial degree
of freedom of every pointer of two subsystems have $x$ and $y$ directional
zero mean Gaussian distribution, and initially there have no any correlation
between them. After taking the two sequential weak measurements with
projection operators where complementary each other, and followed
by a strong measurement on another projection operator where complementary
to the last, the weak average equal to the expectation values of products
of annihilation operators (can be defined in terms of position and
momentum operator) of four Gaussian pointer states. Thus, the real
and imaginary parts of corresponding matrix elements can be found
by calculating the joint positions and momentum shifts of the final
pointer state. Here, we have to mention that previous two sequential
weak measurements caused a spatial shifts on different directions
of the pointer, respectively. 

The rest of the paper is organized as follows: we briefly review the
basic concepts of direct measurement of a quantum state by using postselected
weak measurement based state tomography technique in Section. \ref{sec:Direc}.
In Section. \ref{sec:The-methods-}, we give the details of two methods
to determine the matrix elements of a two photon entangled system,
separately, and take comparison between them and discuss their feasibility.
We give the conclusion to our study in Section. \ref{sec:4}.

\section{\label{sec:Direc} Direct measurement of a state via weak measurement }

From the quantum mechanics we know that the two dimensional photon
polarization state $\vert\psi\rangle$ in Hilbert space can be expressed
in the $\mathcal{A}=\{\vert H\rangle,\vert V\rangle\}$ basis as 
\begin{equation}
\vert\psi\rangle=\sum_{i}c_{i}\vert i\rangle,\ \ \ \ \ \ \!i\in(H,V)\label{eq:eq2}
\end{equation}
where $c_{i}=\langle i\vert\psi\rangle$ is the probability amplitude.
The weak value of projection operator $\text{\ensuremath{\pi_{i}}}=\vert i\rangle\langle i\vert$
with the preselected and postselected states, $\vert\psi\rangle$
and $\vert\alpha\rangle$, is defined as 
\begin{equation}
\langle\pi_{i}\rangle_{\alpha}^{w}=\frac{\langle\alpha\vert\pi_{i}\vert\psi\rangle}{\langle\alpha\vert\psi\rangle}=\frac{1}{\nu}c_{i}.\label{eq:eq4}
\end{equation}
Thus, it is evident that the probability amplitude $c_{i}$ of unknown
state $\vert\psi\rangle$ is directly related with the weak value
of projection operator $\pi_{i}$, and the state vector $\vert\psi\rangle$
can be re-expressed as 
\begin{equation}
\vert\psi\rangle=\sum_{i}\nu\langle\pi_{i}\rangle_{\alpha}^{w}\vert i\rangle.\ \ \ \ \ \ \alpha\in(D,A)\label{eq:eq3}
\end{equation}
Here, $\alpha=D,H$ is the element in $\mathcal{B}=\{\vert D\rangle=\frac{1}{\sqrt{2}}(\vert H\rangle+\vert V\rangle),\vert A\rangle=\frac{1}{\sqrt{2}}(\vert H\rangle-\vert V\rangle)\}$
diagonal and anti-diagonal basis, and $\nu=\frac{\langle\alpha\vert\psi\rangle}{\langle\alpha\vert i\rangle}$
is independent of $i$ and can be determined by the normalization
condition. Since the real and imaginary parts of weak value $\langle\pi_{i}\rangle_{\alpha}^{w}$
can be found simultaneously\citep{PhysRevA.100.032119}, the unknown
state vector $\vert\psi\rangle$ can be reconstruct by optical experiments.
The most important part of this reconstruction technique is the choice
of postselection, and from the Eq.(\ref{eq:eq3}) we can know that
to determine the unknown pure state vector $\vert\psi\rangle$, we
can scan only on definite $\vert\alpha\rangle$ in $\text{\ensuremath{\mathcal{B}}}$
basis at postselection process. However, if we want to reconstruct
the density matrix of a two dimensional unknown state by using weak
measurement technique, we have to take scan through all elements in
both $\mathcal{A}$ and $\mathcal{B}$ bases since its unknown parameters
more than the corresponding pure state. The reconstruction of density
matrix of two dimensional system had been studied experimentally in
Refs.\citep{aBoyd2013,Thekkadath2016}. We have to mention that the
two bases $\mathcal{A}$ and $\mathcal{B}$ are mutually unbiased
for all basis $\vert i\rangle$ in $\mathcal{A}$ and all basis $\vert\alpha\rangle$
in $\mathcal{B}$ in two dimensional Hilbert space, i.g. $\vert\langle i\vert\alpha\rangle\vert^{2}=\frac{1}{2}$. 

\section{\label{sec:The-methods-} The methods of direct measurement of density
operator of an entangled quantum state}

Let us consider a system consisting of two subsystems, and designate
the corresponding state vector as
\begin{equation}
\vert\Psi\rangle=\sum_{ij}C_{ij}\vert i\rangle_{1}\otimes\vert j\rangle_{2}=\sum_{i,j}C_{ij}\vert ij\rangle,\label{eq:e4}
\end{equation}
where $i,j\in(H,V)$, and $C_{ij}=\langle ij\vert\Psi\rangle$ is
complex probability amplitude and $\vert\rangle_{1}$ and $\vert\rangle_{2}$
represent to subsystem one and subsystem two, respectively. To reconstruct
the unknown pure state $\vert\Psi\rangle$, we have to find the corresponding
amplitudes $C_{ij}$ and this task is not very easy as single two
dimensional pure state case. However, recently the Guo-Guang Can et
al.\citep{Pan2019} successfully accomplished this task by using modular
value in stead of weak value of joint projection operators of two
subsystems. In general, the state of a quantum system is characterized
by density matrix and up to now the determination of density matrix
of a two photon entangled state has not been investigated explicitly
yet. The matrix elements of an entangled state described by $\rho$
in $\mathcal{A}^{\prime}=\{\text{\ensuremath{\vert HH\rangle,\vert HV\rangle,\vert VH\rangle,\vert VV\rangle}}\}$
basis of two subsystems is given by 
\begin{align}
\rho= & \text{\ensuremath{\vert\Psi\rangle\langle\Psi\vert=\sum_{ji,kl}C_{ij}C_{kl}^{\ast}\vert ij\rangle\langle kl\vert=\sum_{ij,kl}\rho_{ij,kl}\vert ij\rangle\langle kl\vert}}\nonumber \\
= & \left(\begin{array}{cccc}
\rho_{HH,HH} & \rho_{HV,HH} & \rho_{VH,HH} & \rho_{VV,HH}\\
\rho_{HH,HV} & \rho_{HV,HV} & \rho_{VH,HV} & \rho_{VV,HV}\\
\rho_{HH,VH} & \rho_{HV,VH} & \rho_{VH,VH} & \rho_{VV,VH}\\
\rho_{HH,VV} & \rho_{HV,VV} & \rho_{VH,VV} & \rho_{VV,VV}
\end{array}\right).\label{eq:eq5}
\end{align}
where $\rho_{ij,kl}=\langle ij\vert\rho\vert kl\rangle$ is matrix
element of $\rho$ and a complex number, and $i,j,k,l\in(H,V)$. Thus,
to find the complex matrix elements of an entangled state $\rho$,
we have to find the real and imaginary parts of each elements, $\rho_{ij,kl}$,
respectively. Next we will study this problem with two different methods.

\subsection{Method one: Based on modular value scheme }

As mentioned in Section. \ref{sec:Direc}, the weak value of projection
operator $\text{\ensuremath{\pi_{i}=\vert i\rangle\langle i\vert}}$
under the density operator $\rho$ with postselected state $\vert\alpha\rangle$
is defined as 
\begin{equation}
\text{\ensuremath{\langle\pi_{i}\rangle_{\alpha}^{w}=\frac{\langle\alpha\vert\pi_{i}\rho\vert\alpha\rangle}{\langle\alpha\vert\rho\vert\alpha\rangle}}}\label{eq:eq6}
\end{equation}
 Furthermore, If we want to measure the joint projection operators
of two subsystems, $\pi_{i}^{1}\pi_{j}^{2}=\vert ij\rangle\langle ij\vert$
, where $\pi_{i}^{1}=\vert i\rangle\langle i\vert$ and $\pi_{j}^{2}=\vert j\rangle\langle j\vert$
are represent the projection operators of subsystem one and two, then
the corresponding weak value of $\pi_{i}^{1}\pi_{j}^{2}$ under the
density operator $\text{\ensuremath{\rho}}$ with postselected state
$\text{\ensuremath{\vert\alpha\beta\rangle=\vert\alpha\rangle^{1}\vert\beta\rangle^{2}}}$
can be written as 
\begin{equation}
\text{\ensuremath{\langle\pi_{i}^{1}\pi_{j}^{2}\rangle_{\alpha\beta}^{w}=\frac{\langle\alpha\beta\vert\pi_{i}^{1}\pi_{j}^{2}\rho\vert\alpha\beta\rangle}{\langle\alpha\beta\vert\rho\vert\alpha\beta\rangle}.}}\label{eq:Eq7}
\end{equation}
Here, $i,j,k,l\in(V,H)$ is in $\mathcal{A}^{\prime}$ basis and $\alpha,\beta\in(D,A)$
is in $\mathcal{B}^{\prime}=\{\text{\ensuremath{\vert DD\rangle,\vert DA\rangle,\vert AD\rangle,\vert AA\rangle}}\}$
basis, respectively. 

By using the definition of weak value of joint operators, every matrix
element of $\rho$ which is written in Eq.(\ref{eq:eq5}) can be expressed
in terms of the weak value of joint project operator $\pi_{i}^{1}\pi_{j}^{2}$
in $\mathcal{A}^{\prime}$ basis as

\begin{align}
\rho_{ij,kl} & =\langle ij\vert\rho\vert kl\rangle=\sum_{\alpha\beta}p_{\alpha\beta}\frac{\langle\alpha\beta\vert kl\rangle}{\langle\alpha\beta\vert ij\rangle}\langle\pi_{i}^{1}\pi_{j}^{2}\rangle_{\alpha\beta}^{w},\label{eq:Eq8}
\end{align}
where $p_{\alpha\beta}=\langle\alpha\beta\vert\rho\vert\alpha\beta\rangle$
is the probability to find the system in postselected state$\vert\alpha\beta\rangle$,
$\frac{\langle\alpha\beta\vert kl\rangle}{\langle\alpha\beta\vert ij\rangle}$
is independent to the above summation and can be determined by using
normalization condition. Thus, if we take weak measurement on joint
projection operators $\pi_{i}^{1}\pi_{j}^{2}$ in $\mathcal{A^{\prime}}$
basis following take strong measurement on all elements in $\mathcal{B^{\prime}}$
basis of both subsystems, respectively, then can get the value of
every complex elements of density matrix $\rho$. Furthermore, we
can define the density matrix $\rho$ of an entangled state in $\mathcal{B}^{\prime}$
as well, and the expressions of its matrix elements can be written
as 
\begin{align}
\rho_{\alpha\beta,\alpha^{\prime}\beta^{\prime}} & =\!\!\langle\beta\alpha\vert\rho\vert\alpha^{\prime}\beta^{\prime}\rangle\!\!=\!\!\sum_{ij}p_{\alpha^{\prime}\beta^{\prime}}\frac{\langle\alpha\beta\vert ij\rangle}{\langle\alpha^{\prime}\beta^{\prime}\vert ij\rangle}\langle\pi_{i}^{1}\pi_{j}^{2}\rangle_{\alpha^{\prime}\beta^{\prime}}^{w},\!\!\label{eq:Eq.9}
\end{align}
 where $\vert\alpha^{\prime}\beta^{\prime}\rangle$ also belong to
the $\mathcal{B}^{\prime}$ basis too, i.e., $\alpha^{\prime},\beta^{\prime}\in(D,A)$,
and $p_{\alpha^{\prime}\beta^{\prime}}=\langle\beta^{\prime}\alpha^{\prime}\vert\rho\vert\alpha^{\prime}\beta^{\prime}\rangle$
is the probability of success for postselection of $\vert\alpha^{\prime}\beta^{\prime}\rangle$
basis. As shown in Eq. (\ref{eq:Eq8}) and Eq. (\ref{eq:Eq.9}), to
get the matrix elements $\rho_{ij,kl}$ ( $\rho_{\alpha\beta,\alpha^{\prime}\beta^{\prime}}$)
we should find the weak values $\langle\pi_{i}^{1}\pi_{j}^{2}\rangle_{\alpha\beta}^{w}$
with consider all elements in bases $\mathcal{B^{\prime}}$( $\mathcal{A}^{\prime}$),
respectively. 

In postselected weak measurement technique, the weak value of nonlocal
joint operators can not be obtained exactly and the efficieny is too
low for entangled state case\citep{Lundeen2005}. However, in recent
study of Guang-Can- Guo \citep{Pan2019}, they showed that the joint
weak values $\text{\ensuremath{\langle\pi_{i}^{1}\pi_{j}\rangle_{\alpha\beta}^{w}}}$
can be found in terms of modular values. In remaining part of this
subsection, we will calculate the weak values $\langle\pi_{i}^{1}\pi_{j}^{2}\rangle_{\alpha\beta}^{w}$
to find the matrix elements of $\rho$. 

The modular value of an observable $\hat{F}$ with pre- and postselected
states, $\vert\psi_{in}\rangle$ and $\vert\psi_{fi}\rangle$ can
be written as\citep{Vaidman2010} 
\begin{equation}
\langle F\rangle_{\psi_{fi}}^{m}=\frac{\langle\psi_{fi}\vert e^{-igF}\vert\psi_{in}\rangle}{\langle\psi_{fi}\vert\psi_{in}\rangle}.\label{eq:Eq9}
\end{equation}
 where $g$ is represent the coupling strength between measured system
and pointer, and the modular value is valid for any weak and strong
coupling cases. If we take $\hat{F}=\hat{\pi}=\vert i\rangle\langle i\vert$
is a projection operator in two dimensional Hilbert space, then 
\begin{align}
\langle\pi\rangle_{\psi_{fi}}^{m} & =\frac{\langle\psi_{fi}\vert(\sum_{i}e^{-ig\lambda_{i}}\text{\ensuremath{\pi_{i}}})\vert\psi_{in}\rangle}{\langle\psi_{fi}\vert\psi_{in}\rangle}\nonumber \\
 & =\frac{\langle\psi_{fi}\vert((1-\pi_{i}+e^{-ig}\pi_{i})\vert\psi_{in}\rangle}{\langle\psi_{fi}\vert\psi_{in}\rangle}\nonumber \\
 & =1+(e^{-ig}-1)\frac{\langle\psi_{fi}\vert\pi_{i}\vert\psi_{in}\rangle}{\langle\psi_{fi}\vert\psi_{in}\rangle}\nonumber \\
 & =1+s\langle\pi_{i}\rangle_{\psi_{fi}}^{w}\label{eq:Eq10}
\end{align}
 where $\text{\ensuremath{\lambda_{i}=0,1}}$ are eigenvalues of projection
operator $\pi_{i}$, and $s=e^{-ig}-1$. 

Furthermore, if we extend our concern to an entangled state composed
of two subsystems, i.e., consider $\vert\Psi\rangle$(Eq.(\ref{eq:eq4}))
as preselection state of the system, then the modular value of projection
operators $\pi_{i}^{1}+\pi_{j}^{2}=\vert i\rangle_{1}\langle i\vert+\vert j\rangle_{2}\langle j\vert$
of total system with postselected state $\vert\alpha\beta\rangle$
can be calculated as 
\begin{align}
\langle\pi_{i}^{1}+\pi_{j}^{2}\rangle_{\alpha\beta}^{m} & =\frac{\langle\beta\alpha\vert e^{-ig(\pi_{i}^{1}+\pi_{j}^{2})}\vert\Psi\rangle}{\langle\alpha\beta\vert\Psi\rangle}\nonumber \\
 & =\frac{\langle\beta\alpha\vert e^{-ig\pi_{i}^{1}}e^{-ig\pi_{j}^{2}}\vert\Psi\rangle}{\langle\alpha\beta\vert\Psi\rangle}\nonumber \\
 & =\frac{\langle\beta\alpha\vert(1+s\pi_{i}^{1})(1+s\pi_{j}^{2})\vert\Psi\rangle}{\langle\alpha\beta\vert\Psi\rangle}\nonumber \\
 & =\frac{\langle\beta\alpha\vert(1+s\pi_{i}^{1})(1+s\pi_{j}^{2})\vert\Psi\rangle}{\langle\alpha\beta\vert\Psi\rangle}\nonumber \\
 & =1+s\langle\pi_{i}^{1}\rangle_{\alpha\beta}^{w}+s\langle\pi_{j}^{2}\rangle_{\alpha\beta}^{w}+s^{2}\langle\pi_{i}^{1}\pi_{j}^{2}\rangle_{\alpha\beta}^{w}\nonumber \\
 & =-1+\langle\pi_{i}^{1}\rangle_{\alpha\beta}^{m}+\langle\pi_{j}^{2}\rangle_{\alpha\beta}^{m}+s^{2}\langle\pi_{i}^{1}\pi_{j}^{2}\rangle_{\alpha\beta}^{w}.\label{eq:Eq11}
\end{align}
 In the last line of above expression we use relations between weak
value and modular value (see Eq. (\ref{eq:Eq10})). By taking the
modular value of projection operator $\text{\ensuremath{\pi_{i}=\vert i\rangle\langle i\vert}}$
which is given in Eq.(\ref{eq:Eq10}) into account, from this above
equation we can read the weak value $\langle\pi_{i}^{1}\pi_{j}^{2}\rangle_{\alpha\beta}^{w}$
as
\begin{equation}
\langle\pi_{i}^{1}\pi_{j}^{2}\rangle_{\alpha\beta}^{w}=s^{-2}[\langle\pi_{i}^{1}+\pi_{j}^{2}\rangle_{\alpha\beta}^{m}-\langle\pi_{i}^{1}\rangle_{\alpha\beta}^{m}-\langle\pi_{j}^{2}\rangle_{\alpha\beta}^{m}+1].\label{eq:Eq.12}
\end{equation}
 From this theoretical result we can deduce that if we can measure
the modular values of projection operators $\pi_{i}^{1},\pi_{j}^{2}$
and $\pi_{i}^{1}+\pi_{j}^{2}$, respectively, the weak value of joint
operators $\pi_{i}^{1}\pi_{j}^{2}$ can be found easily, after that
we can determine the matrix elements $\rho_{ij,kl}$ and $\rho_{\alpha\beta,\alpha^{\prime}\beta^{\prime}}$
of density matrix $\rho$ by using Eq.(\ref{eq:Eq8}) and Eq.(\ref{eq:Eq.9}). 

As studied in Ref.\citep{Pan2019}, after a two photon entangled state
generated by nonlinear optical devices, during the propagation in
space (interferometer for example) the two photons entangled in their
polarization degrees of freedom and paths degrees of freedom, respectively.
We take the paths degrees of freedom( $\vert\uparrow\rangle$ and
$\vert\downarrow\rangle$) as pointer, and polarization degrees of
freedom ($\vert H\rangle$ and $\vert V\rangle$) take as measured
system, respectively. Suppose that initially both paths and polarization
degrees of two subsystems are entangled but there is no any entanglement
between these two degrees of freedoms. Thus, the initial state of
the total system can be expressed as
\begin{equation}
\vert\Psi_{ms}\rangle=\vert\varphi\rangle\otimes\vert\text{\ensuremath{\Psi}}\rangle,\label{eq:eq13}
\end{equation}
where, $\vert\Psi\rangle$ is given in Eq. (\ref{eq:e4}) and 
\begin{equation}
\vert\varphi\rangle=\mu\vert\uparrow\downarrow\rangle+\eta\vert\downarrow\uparrow\rangle,\ \ \ \ \ \vert\mu\vert^{2}+\vert\eta\vert^{2}=1\label{eq:eq14}
\end{equation}
is correspond to paths degree of freedom of two component systems.
Here, $\vert\uparrow\downarrow\rangle$ represent the first photon
in the $\uparrow$ path and second photon in $\downarrow$ path, respectively.
To get modular value of $\pi_{i}^{1},\pi_{j}^{2}$ and $\pi_{i}^{1}+\pi_{j}^{2}$
, in Ref.\citep{Pan2019} they introduced the three interaction Hamiltonians
between two composed pointer state and measured system as 

\begin{align}
H_{1} & =g\delta(t-t_{0})(\pi_{\downarrow}^{1}\pi_{i}^{1}+\pi_{\uparrow}^{2}\pi_{j}^{2}),\label{eq:EQ15}\\
H_{2} & =g\delta(t-t_{0})\pi_{\downarrow}^{1}\pi_{i}^{1},\label{eq:EQ16}\\
H_{3} & =g\delta(t-t_{0})\pi_{\uparrow}^{2}\pi_{j}^{2}.\label{eq:eq17}
\end{align}
 Here, $\pi_{\downarrow}^{1}=\vert\downarrow\rangle\langle\downarrow\vert$
and $\pi_{\uparrow}^{2}=\vert\uparrow\rangle\langle\uparrow\vert$
are represent the projection operators of paths degree of freedom
of two subsystems, respectively. 

If we consider the intrinsic properties of projection operators of
paths degrees of freedom, the evolution operators corresponding to
above interaction Hamiltonians becomes as\begin{subequations}

\begin{align}
U_{1} & =\exp[-\frac{i}{\hbar}\int Hd\tau]=e^{-ig(\pi_{\downarrow}^{1}\pi_{i}^{1}+\pi_{\uparrow}^{2}\pi_{j}^{2})}\nonumber \\
 & =[1+(e^{-ig\pi_{i}^{1}}-1)\pi_{\downarrow}^{1}][1+(e^{-ig\pi_{j}^{2}}-1)\pi_{\uparrow}^{2}]\nonumber \\
 & =(e^{-ig\pi_{i}^{1}}-1)\pi_{\downarrow}^{1}+(e^{-ig\pi_{j}^{2}}-1)\pi_{\uparrow}^{2}+1\nonumber \\
 & +[e^{-ig(\pi_{i}^{1}+\pi_{j}^{2})}+1-e^{-ig\pi_{i}^{1}}-e^{-ig\pi_{l}^{2}}]\pi_{\downarrow}^{1}\pi_{\uparrow}^{2},\label{eq:19}\\
U_{2} & =e^{-ig\pi_{\downarrow}^{1}\pi_{j}^{2}}=1+(e^{-ig\pi_{i}^{1}}-1)\pi_{\downarrow}^{1},\label{eq:20}\\
U_{3} & =e^{-ig\pi_{\uparrow}^{2}\pi_{j}^{2}}=1+(e^{-ig\pi_{j}^{2}}-1)\pi_{\uparrow}^{2},\label{eq:21}
\end{align}
\end{subequations}respectively. In above calculations we use the
formula $e^{\theta\hat{F}}=\sum_{n}e^{\theta\lambda_{n}}\vert\phi_{n}\rangle\langle\phi_{n}\vert$
of operator $\hat{F}$ with $\text{\ensuremath{\hat{F}\vert\phi_{n}\rangle=\lambda_{n}\vert\phi_{n}\rangle}}$.

Start from the initial state of the total system, Eq.(\ref{eq:eq13}),
and take the above time evolution operators ( see Eqs.(\ref{eq:19}-\ref{eq:21}))
and post-selection onto $\vert\alpha\beta\rangle$ in basis $\mathcal{B}^{\prime}$
into account, the final states of the pointers can be obtained as
\begin{align}
\vert\Phi_{1}\rangle & =\mathcal{N}_{1}[\eta\langle\pi_{i}^{1}+\pi_{j}^{2}\rangle_{\alpha\beta}^{m}\vert\downarrow\uparrow\rangle+\mu\vert\uparrow\downarrow\rangle],
\end{align}
\begin{align}
\vert\Phi_{2}\rangle & =\mathcal{N}_{2}[\eta\langle\pi_{i}^{1}\rangle_{\alpha\beta}^{m}\vert\downarrow\uparrow\rangle+\mu\vert\uparrow\downarrow\rangle],\label{eq:Eq25}
\end{align}
 and 
\begin{align}
\vert\Phi_{2}\rangle & =\mathcal{N}_{3}[\eta\langle\pi_{j}^{2}\rangle_{\alpha\beta}^{m}\vert\downarrow\uparrow\rangle+\mu\vert\uparrow\downarrow\rangle],\label{eq:Eq26}
\end{align}
 where $\mathcal{N}_{1}=[\vert\mu\vert^{2}+\vert\eta\langle\pi_{i}^{1}+\pi_{j}^{2}\rangle_{\alpha\beta}^{m}\vert^{2}]^{-\frac{1}{2}}$,$\mathcal{N}_{1}=[\vert\mu\vert^{2}+\vert\eta\langle\pi_{i}^{1}\rangle_{\alpha\beta}^{m}\vert^{2}]^{-\frac{1}{2}}$
and $\mathcal{N}_{3}=\mathcal{N}_{1}=[\vert\mu\vert^{2}+\vert\eta\langle\pi_{j}^{2}\rangle_{\alpha\beta}^{m}\vert^{2}]^{-\frac{1}{2}}$
are normalization coefficients, respectively. 

If we project these above final states of the pointer onto $\vert\varphi_{1}\rangle=\frac{1}{\sqrt{2}}(\vert\uparrow\rangle+\vert\downarrow\rangle)\otimes\frac{1}{\sqrt{2}}(\vert\uparrow\rangle+\vert\downarrow\rangle)$
and $\vert\varphi_{2}\rangle=\frac{1}{\sqrt{2}}(\vert\uparrow\rangle+i\vert\downarrow\rangle)\otimes\frac{1}{\sqrt{2}}(\vert\uparrow\rangle+i\vert\downarrow\rangle)$,
respectively, the probabilities to find the final states $\vert\Phi_{1}\rangle,$$\vert\Phi_{2}\rangle$
and $\vert\Phi_{3}\rangle$ on $\vert\varphi_{1}\rangle$ and $\vert\varphi_{2}\rangle$
are 

\begin{subequations}

\begin{align}
P_{1} & =\vert\langle\varphi_{1}\vert\Phi_{1}\rangle\vert^{2}\\
 & =\frac{\vert\mathcal{N}_{1}\vert^{2}}{2}\{\vert\mu\vert^{2}+\vert\eta\vert^{2}\vert\langle\pi_{i}^{1}+\pi_{j}^{2}\rangle_{\alpha\beta}^{m}\vert^{2}+2\Re[\mu^{\ast}\eta\langle\pi_{i}^{1}+\pi_{j}^{2}\rangle_{\alpha\beta}^{m}]\}\\
P_{2} & =\vert\langle\varphi_{2}\vert\Phi_{1}\rangle\vert^{2}\\
 & =\frac{\vert\mathcal{N}_{1}\vert^{2}}{2}\{\vert\mu\vert^{2}+\vert\eta\vert^{2}\vert\langle\pi_{i}^{1}+\pi_{j}^{2}\rangle_{\alpha\beta}^{m}\vert^{2}+2\Im[\mu^{\ast}\eta\langle\pi_{i}^{1}+\pi_{j}^{2}\rangle_{\alpha\beta}^{m}]\}\\
P_{3} & =\vert\langle\varphi_{1}\vert\Phi_{2}\rangle\vert^{2}\\
 & =\frac{\vert\mathcal{N}_{2}\vert^{2}}{2}\{\vert\mu\vert^{2}+\vert\eta\vert^{2}\vert\langle\pi_{i}^{1}\rangle_{\alpha\beta}^{m}\vert^{2}+2\Re[\mu^{\ast}\eta\langle\pi_{i}^{1}\rangle_{\alpha\beta}^{m}]\}\\
P_{5} & =\vert\langle\varphi_{2}\vert\Phi_{2}\rangle\vert^{2}\\
 & =\frac{\vert\mathcal{N}_{2}\vert^{2}}{2}\{\vert\mu\vert^{2}+\vert\eta\vert^{2}\vert\langle\pi_{i}^{1}\rangle_{\alpha\beta}^{m}\vert^{2}+2\Im[\mu^{\ast}\eta\langle\pi_{i}^{1}\rangle_{\alpha\beta}^{m}]\}
\end{align}
and 
\begin{align}
P_{5} & =\vert\langle\varphi_{1}\vert\Phi_{3}\rangle\vert^{2}\nonumber \\
 & =\frac{\vert\mathcal{N}_{3}\vert^{2}}{2}\{\vert\mu\vert^{2}+\vert\eta\vert^{2}\vert\langle\pi_{j}^{1}\rangle_{\alpha\beta}^{m}\vert^{2}+2\Re[\mu^{\ast}\eta\langle\pi_{j}^{1}\rangle_{\alpha\beta}^{m}]\}
\end{align}
\begin{align}
P_{6} & =\vert\langle\varphi_{2}\vert\Phi_{3}\rangle\vert^{2}\nonumber \\
 & =\frac{\vert\mathcal{N}_{3}\vert^{2}}{2}\{\vert\mu\vert^{2}+\vert\eta\vert^{2}\vert\langle\pi_{j}^{1}\rangle_{\alpha\beta}^{m}\vert^{2}+2\Im[\mu^{\ast}\eta\langle\pi_{j}^{1}\rangle_{\alpha\beta}^{m}]\},
\end{align}
 \end{subequations}respectively. If we assume that initially the
probability of first photon in path $\text{\ensuremath{\downarrow}}$
and second photon in path $\uparrow$ is smaller than the probability
of first photon in path $\uparrow$ and second photon in path $\downarrow$
, i.e, $\vert\eta\vert^{2}\ll1$ , then 

\begin{subequations}

\begin{align}
\mathcal{P}_{1} & \approx\mu\Re[\langle\pi_{i}^{1}+\pi_{j}^{2}\rangle_{\alpha\beta}^{m}]+\frac{1}{2},\\
\mathcal{P}_{2} & \approx\mu\Im[\langle\pi_{i}^{1}+\pi_{j}^{2}\rangle_{\alpha\beta}^{m}]+\frac{1}{2},\\
\mathcal{P}_{3} & =\mu\Re[\langle\pi_{i}^{1}\rangle_{\alpha\beta}^{m}]+\frac{1}{2},\\
\mathcal{P}_{4} & =\mu\Im[\langle\pi_{i}^{1}\rangle_{\alpha\beta}^{m}]+\frac{1}{2},\\
\mathcal{P}_{5} & =\mu\Re[\langle\pi_{j}^{2}\rangle_{\alpha\beta}^{m}]+\frac{1}{2},\\
\mathcal{P}_{6} & =\mu\Im[\langle\pi_{j}^{2}\rangle_{\alpha\beta}^{m}]+\frac{1}{2}.
\end{align}

\end{subequations}

These probabilities can be determine by the detectors in the Lab,
then we can find the modular values $\pi_{i}^{1},\pi_{j}^{2}$ and
$\pi_{i}^{1}+\pi_{j}^{2}$. Finally, the real and imaginary parts
of weak value of $\pi_{i}^{1}\pi_{j}^{2}$ ( Eq.(\ref{eq:Eq.12}))
can be written as
\begin{align}
\Re[\langle\pi_{i}^{1}\pi_{j}^{2}\rangle_{\alpha\beta}^{w}] & =s^{-2}\eta^{-1}[\mathcal{P}_{1}-\mathcal{P}_{3}-\mathcal{P}_{5}+\eta+\frac{1}{2}]
\end{align}
 and 
\begin{equation}
\Im[\langle\pi_{i}^{1}\pi_{j}^{2}\rangle_{\alpha\beta}]=s^{-2}\eta^{-1}[\mathcal{P}_{2}-\mathcal{P}_{4}-\mathcal{P}_{6}+\frac{1}{2}],
\end{equation}
respectively. With these processes finally we can determine the matrix
elements of density operator by using Eq.(\ref{eq:Eq8}) and Eq.(\ref{eq:Eq.9}),
respectively. 

In Ref.\citep{Pan2019}, they investigated the direct measurement
method of a pure two photon polarization entangled state theoretically
and experimentally. In their work to get the complex amplitude $C_{ij}$
in Eq.(\ref{eq:eq4}) we only need to scan a definite element of $\mathcal{B}^{\prime}$
basis, i.e.
\begin{equation}
C_{ij}=\chi\langle\pi_{i}^{1}\pi_{j}^{2}\rangle_{DD}^{w}\label{eq:Eq38}
\end{equation}
where $\chi=\frac{\langle DD\vert\Psi\rangle}{\langle DD\vert ij\rangle}$
is independent of $\vert ij\rangle$ and can be obtained by normalization
condition. However, to reconstruct the density matrix of two component
entangled state in $\mathcal{A}^{\prime}$ basis, we need more strong
measurement steps in $\mathcal{B}^{\prime}$ basis to determine every
matrix elements. For example, if we want to get the matrix element
$\rho_{HH,HV}$, according to Eq.(\ref{eq:Eq8}) we should find all
weak values of $\pi_{i}^{1}\pi_{j}^{2}$ with postselection states
in $\mathcal{B}^{\prime}$ basis. i.e. 
\begin{align}
\rho_{HH,HV} & =p_{DD}\langle\pi_{H}^{1}\pi_{V}^{2}\rangle_{DD}^{w}+p_{DA}\langle\pi_{H}^{1}\pi_{V}^{2}\rangle_{DA}^{w}\nonumber \\
 & +p_{AD}\langle\pi_{H}^{1}\pi_{V}^{2}\rangle_{AD}^{w}+p_{AA}\langle\pi_{H}^{1}\pi_{V}^{2}\rangle_{AA}^{w}.\label{eq:eq39}
\end{align}
On the other hand, if we want to get the matrix elements $\rho_{\alpha\beta,DA}$
for example, according to Eq.(\ref{eq:Eq.9}) we should find all weak
values of $\pi_{i}^{1}\pi_{j}^{2}$ in $\mathcal{A^{\prime}}$ basis
with definite postselection state $\vert DA\rangle$ in $\mathcal{B}^{\prime}$
basis,i.e.
\begin{align}
\rho_{DD,DA} & =2p_{DA}[\langle\pi_{H}^{1}\pi_{H}^{2}\rangle_{DA}^{w}+\langle\pi_{V}^{1}\pi_{H}^{2}\rangle_{DA}^{w}-\frac{1}{2}].\label{eq:eQ41}
\end{align}
 Here, we use the relation of weak value of projection operators $\pi_{i}^{1}\pi_{j}^{2}$
in whole Hilbert space of our system, i.e., 
\begin{equation}
\sum_{ij}\langle\pi_{i}^{1}\pi_{j}^{2}\rangle_{\alpha\beta}^{w}=1.\label{eq:Eq40}
\end{equation}
From the above examples we can deduce that the determination of matrix
elements $\rho_{ij,kl}$ in $\mathcal{A^{\prime}}$ basis need more
experimental steps than the matrix elements $\rho_{\alpha\beta,\alpha^{\prime}\beta^{\prime}}$
in $\mathcal{B^{\prime}}$ basis.

Thus, if we want to get the matrix elements of density operator $\rho$
in $\mathcal{A}^{\prime}$ basis, it could be realized in experiment
based on the the experimental setup of Guang-Can- Guo \citep{Pan2019}
by extend the chooses of the post-selection state to all elements
in $\mathcal{B}^{\prime}$ basis rather than only scan on one definite
element in $\mathcal{B}^{\prime}$.

\subsection{Method Two: Based on three sequential measurements \citep{Lundeen2012}scheme}

As Lundeen and his co-workers \citep{Lundeen2012} studied, the matrix
elements of a quantum system can be obtained by considering the weak
measurement of an observable composed of three incompatible projection
operator:
\begin{equation}
\Pi_{ij}=\pi_{j}\pi_{D}\pi_{i}.\label{eq:eq43}
\end{equation}
Here, $\pi_{i}=\vert i\rangle\langle i\vert,\pi_{j}=\vert j\rangle\langle j\vert$
with $i,j\in(H,V)$ in $\mathcal{A}$ basis, and $\pi_{D}=\vert D\rangle\langle D\vert$
where $\vert D\rangle=\frac{1}{\sqrt{2}}(\vert H\rangle+\vert V\rangle)$
is element in $\mathcal{B}$ basis. The basis vectors in $\mathcal{A}$
and $\,\mathcal{B}$ are maximally incompatible, and $\langle i\vert D\rangle=\langle j\vert D\rangle=\frac{1}{\sqrt{2}}$.
The matrix elements $\text{\ensuremath{\rho_{ij}}}$ of unknown density
operator $\rho$ can be found as 
\begin{equation}
\rho_{ij}=2\langle\Pi_{ij}\rangle_{s}=2Tr_{s}[\pi_{j}\pi_{D}\pi_{i}\rho].\label{eq:Eq.44}
\end{equation}
 Since $\Pi_{ij}$ is non-Hermitian, generally the weak average $\langle\Pi_{ij}\rangle_{s}$
is a complex number. Thus, according to the Eq.(\ref{eq:Eq.44}) we
can get the complex density matrix elements of $\rho$ if one can
find the $\langle\Pi_{ij}\rangle_{s}$. In the recent work of Lundeen
and his co-workers, they investigated their proposal which introduced
in Ref.\citep{Lundeen2012}, and experimentally reconstruct the density
matrix elements of pure and mixed states of 2-dimensional system\citep{Thekkadath2016}.
In this study as expansion of their work \citep{Thekkadath2016},
we will study how to determine the matrix elements of two photon entangled
state. 

For an entangled state composed of two subsystems, the observable
defined in Eq.(\ref{eq:eq43}), can be redefined as 
\begin{equation}
\Pi_{ij,kl}=\pi_{kl}\pi_{\alpha\beta}\pi_{ij}
\end{equation}
where $\pi_{ij}=\pi_{i}^{1}\pi_{j}^{2}=\vert i\rangle^{1}\langle i\vert\otimes\vert j\rangle^{2}\langle j\vert=\vert ij\rangle\langle ij\vert,\pi_{kl}=\pi_{k}^{1}\pi_{l}^{2}=\vert k\rangle^{1}\langle k\vert\otimes\vert l\rangle^{2}\langle l\vert=\vert kl\rangle\langle kl\vert$
with $i,j,k,l\in(H,V)$ in $\mathcal{A^{\prime}}$ basis, and $\pi_{\alpha\beta}=\pi_{\alpha}^{1}\pi_{\beta}^{2}=\vert\alpha\rangle^{1}\langle\alpha\vert\otimes\vert\beta\rangle^{2}\langle\beta\vert=\vert\alpha\beta\rangle\langle\alpha\beta\vert$
with $\alpha,\beta\in(D,A)$ in $\mathcal{B^{\prime}}$ basis. The
basis vectors in $\mathcal{A^{\prime}}$ and $\mathcal{B^{\prime}}$are
maximally incompatible, and $\langle ij\vert\alpha\beta\rangle=\langle kl\vert\alpha\beta\rangle=\frac{1}{2}$.
The matrix elements $\text{\ensuremath{\rho_{ij,kl}}}$ of unknown
density operator $\rho$ of an entangled state can be found as 
\begin{equation}
\rho_{ij,kl}=4\langle\Pi_{ij,kl}\rangle_{s}=4Tr_{s}[\pi_{kl}\pi_{\alpha\beta}^{o}\pi_{ij}\rho].\label{eq:Eq.44-1}
\end{equation}
where $\pi_{\alpha\beta}^{o}$ represent the two composed project
operator with definite value of $\alpha$ and $\beta$ in $\mathcal{B}^{\prime}$
basis. Here, we will only consider the $\alpha=\beta=D$ case with
following the method of Lundeen\citep{Lundeen2012}, but other cases
such as $\alpha=\beta=A$ also can be used to find the matrix elements
with similar processes described in this study. To find the matrix
elements $\rho_{ij,kl}$, we have to find the value of $Tr_{s}[\pi_{kl}\pi_{\alpha\beta}^{o}\pi_{ij}\rho]$,
and remaining part of this subsection we will study this problem. 

We assume that the initial state of total system is 
\[
\vert\Omega\rangle=\vert\text{\ensuremath{\varPhi}}\rangle\langle\text{\ensuremath{\varPhi}}\vert\otimes\rho,
\]
 where the initial state of the pointer $\Phi(r_{1},r_{2})$ is composed
by two Gaussian beams of two photons which have $x$ and $y$ transverse
spatial distributions separately, i.e. 
\begin{equation}
\langle r\vert\Phi\rangle=\varphi_{1}(x_{1},y_{1})\varphi_{2}(x_{2},y_{2})
\end{equation}
 where 
\begin{align}
\varphi_{1}(x_{1},y_{1}) & =\left(\frac{1}{2\pi\sigma_{x_{1}}\sigma_{y_{1}}}\right)^{\frac{1}{2}}\exp\left(\frac{x_{1}^{2}}{4\sigma_{x_{1}}^{2}}\right)\exp\left(\frac{y_{1}^{2}}{4\sigma_{y_{1}}^{2}}\right)
\end{align}
 and 
\begin{equation}
\varphi_{2}(x_{2},y_{2})=\left(\frac{1}{2\pi\sigma_{x_{2}}\sigma_{y_{2}}}\right)^{\frac{1}{2}}\exp\left(\frac{x_{2}^{2}}{4\sigma_{x_{2}}^{2}}\right))\exp\left(\frac{y_{2}^{2}}{4\sigma_{y_{2}}^{2}}\right).
\end{equation}
 are represent the spatial distributions of first and second photons,
respectively. $\rho$ is given in Eq.(\ref{eq:eq5}), and considered
as measured system. 

We assume that the interaction Hamiltonian between each pointer and
measured systems are 
\begin{equation}
H=H_{1}+H_{2}+H_{3}+H_{4},
\end{equation}
with 
\begin{equation}
H_{1}=g_{1}\pi_{i}^{1}x_{1},H_{2}=g_{2}\pi_{j}^{2}x_{2},\ \ \ \ i,j\in(V,H)
\end{equation}
and 
\begin{equation}
H_{3}=g_{3}\pi_{D}^{1}y_{1},H_{4}=g_{4}\pi_{D}^{2}y_{2},
\end{equation}
respectively, and $g_{n}$($n=1,2,3,4$) represent the coupling strength
between the pointer and measuring device, and for simplicity can be
taken them as equal quantity, i.e., $g_{1}=g_{2}=g_{3}=g_{4}=g$.
If we assume that the polarizers represented by the projection operator
$\pi_{ij}$ and $\pi_{DD}$ causing displacement along $x$ and $y$
directions, respectively, we can get the weak average of $\pi_{DD}\pi_{ij}$
by using the method introduced in Refs.\citep{Lundeen2005,Lundeen2012}
as 
\begin{equation}
\langle\pi_{DD}\pi_{ij}\rangle_{s}=\frac{1}{g^{4}}\langle a_{2D}a_{1D}a_{2j}a_{1i}\rangle_{f},\label{eq:eQ53}
\end{equation}
 where
\begin{equation}
a_{1i}=x_{1i}+\iota\frac{2\sigma^{2}}{\hbar}p_{1xi},\ \ \ \ \ a_{2j}=x_{2j}+\iota\frac{2\sigma^{2}}{\hbar}p_{2xj},
\end{equation}
 and 
\begin{equation}
a_{1D}=y_{1D}+\iota\frac{2\sigma^{2}}{\hbar}p_{1yD},\ \ \ \ \ a_{2D}=y_{2D}+\iota\frac{2\sigma^{2}}{\hbar}p_{2yD},
\end{equation}
are represent the annihilation operators of every spatial transversal
components of each photons, respectively, and $\langle\rangle_{f}$
indicate to find the expectation value of variables under the final
state of the pointer state. Here, we have to note that $\pi_{ij}$
and $\pi_{DD}$ are non-commute, but as showed in Ref.\citep{PhysRevA.77.052102}
the Eq.(\ref{eq:eQ53}) still valid for non-commuting observables
if they are measured sequentially as measuring the $\pi_{DD}$ followed
by $\pi_{ij}$. Since last measurement is will be taken over the projection
operators $\pi_{kl}$ are strong, then 
\begin{equation}
Tr_{s}[\pi_{kl}\pi_{\alpha\beta}\pi_{ij}\rho]=\frac{1}{g^{4}}Tr_{s}[\pi_{kl}a_{2D}a_{1D}a_{2j}a_{1i}\rho].
\end{equation}
With these processes we can obtain the matrix elements $\rho_{ij,kl}$
of density operator $\rho$ as

\begin{widetext}

\begin{align}
\rho_{ij,kl} & =4Tr[\pi_{kl}\pi_{DD}\pi_{ij}\rho]=4Tr[\pi_{kl}a_{1D}a_{2D}a_{2j}a_{1i}\rho]\nonumber \\
 & =\frac{4}{g^{4}}\text{\ensuremath{\left\langle \left(y_{1D}+\iota\frac{2\sigma_{y1}^{2}}{\hbar}p_{1yD}\right)\left(y_{2D}+\iota\frac{2\sigma_{y2}^{2}}{\hbar}p_{2yD}\right)\left(x_{2j}+\iota\frac{2\sigma_{x2}^{2}}{\hbar}p_{2xj}\right)\left(x_{1i}+\iota\frac{2\sigma_{x1}^{2}}{\hbar}p_{1xi}\right)\right\rangle _{f}}}\label{eq:eq51}
\end{align}
Then, the real and imaginary parts of the matrix elements of density
operator $\rho$ are 

\begin{align}
\Re[\rho_{ij,kl}] & =\frac{4}{g^{4}}[\langle y_{1D}y_{2D}x_{1i}x_{2j}\rangle_{f}-\frac{\sigma^{2}}{\sigma_{p}^{2}}\langle y_{1D}y_{2D}p_{1xi}p_{2xj}\rangle_{f}-\frac{\sigma^{2}}{\sigma_{p}^{2}}\langle x_{1i}x_{2j}p_{2yD}p_{1yD}\rangle_{f}+\frac{\sigma^{4}}{\sigma_{p}^{4}}\langle p_{1xD}p_{2xj}p_{2yD}p_{1yD}\rangle_{f}\nonumber \\
 & -\text{\ensuremath{\frac{\sigma^{2}}{\sigma_{p}^{2}}\langle y_{1D}p_{2yD}x_{2j}p_{1xi}\rangle_{f}-\frac{\sigma^{2}}{\sigma_{p}^{2}}\langle y_{1D}p_{2yD}x_{1i}p_{2xj}\rangle_{f}-\frac{\sigma^{2}}{\sigma_{p}^{2}}\langle y_{2D}p_{1yD}x_{2j}p_{1xi}\rangle_{f}-\frac{\sigma^{2}}{\sigma_{p}^{2}}\langle y_{2D}p_{1yD}x_{1i}p_{2xj}\rangle_{f}]}}\label{eq:e151}
\end{align}
and

\begin{align}
\Im[\rho_{ij,kl}] & =\frac{4}{g^{4}}\frac{\sigma}{\sigma_{p}}[\langle y_{1D}y_{2D}x_{2j}p_{1xi}\rangle_{f}+\langle y_{1D}y_{2D}x_{1i}p_{2xj}\rangle_{f}+\langle x_{1i}x_{2j}y_{1D}p_{2yD}\rangle_{f}+\langle x_{1i}x_{2j}y_{2D}p_{1yD}\rangle_{f}\nonumber \\
 & -\frac{\text{\ensuremath{\sigma^{2}}}}{\sigma_{p}^{2}}\langle p_{2yD}p_{1yD}x_{2j}p_{1xi}\rangle_{f}-\frac{\text{\ensuremath{\sigma^{2}}}}{\sigma_{p}^{2}}\langle p_{2yD}p_{1yD}x_{1i}p_{2xj}\rangle_{f}-\frac{\text{\ensuremath{\sigma^{2}}}}{\sigma_{p}^{2}}\langle p_{1xi}p_{2xj}y_{1D}p_{2yD}\rangle_{f}-\frac{\text{\ensuremath{\sigma^{2}}}}{\sigma_{p}^{2}}\langle p_{1xi}p_{2xj}y_{2D}p_{1yD}\rangle_{f}]\label{eq:eq52}
\end{align}

\end{widetext}, respectively. Here, for simplicity we assume that
the width of every Gaussian beam is equal to $\text{\ensuremath{\sigma}}$,
i.e., $\sigma_{x1}=\sigma_{y1}=\sigma_{x2}=\sigma_{y2}=\sigma$, and
$\sigma_{p}$ is the momentum space width of the pointer state with
$\sigma\sigma_{p}=\frac{\hbar}{2}$. Based on the experimental results
and methods of Ref.\citep{Thekkadath2016} for read out the real and
imaginary parts of matrix elements of single photon polarization state
by measuring the probabilities of transmitted photons via optical
apparatuses, in the Lab we may also measure the probabilities of entangled
photons transmitted through the final polarizers which represented
by the projection operators $\pi_{kl}=\vert kl\rangle\langle kl\vert$
of two subsystems. In general, these probabilities are functions of
positions and momenta of two photons, i.e, $\mathcal{P}=P(x_{1},y_{1},y_{2},x_{2},p_{1x},p_{2x},p_{2y},p_{2y})$.
Then, the elements of density operator of two entangled photon state
can be reconstructed by determining the expectation values $\langle\rangle_{f}$
in Eq.(\ref{eq:e151}) and Eq.(\ref{eq:eq52}) via $\int ABCD\mathcal{P}d\tau=\langle ABCD\rangle_{f}$,
respectively. 

\section{\label{sec:4}Conclusion and remarks}

In this study we investigated how to reconstruct the unknown density
operator of two component entangled quantum state by using postselected
weak measurement method and three sequential measurements where each
complementary to the last, respectively, and discussed its feasibility
by taking into account the recent related experimental works. The
similarity of these methods is that in each scheme we take the weak
and strong sequential measurements in two bases during the getting
of real and imaginary parts of elements of density operator, respectively,
and the postselection is the key to determine which matrix element
we want to readout from the final state of the pointer state. However,
in second method it is enough to scan over one definite elements in
bases $\mathcal{B}^{\prime}$ but in first method we usually need
to scan all elements in $\mathcal{B}^{\prime}$. Thus, in the same
measurement process the method one may need more resources than method
two. 

Since the Hilbert space of two entangled photon state is larger than
single photon case, during the readout of the matrix elements in the
Lab of a two entangled photon processes we would need more and some
complicated experimental setups and need more resources in method
one rather than method two. However, if we consider the wide applications
of entangled photon states in every field of quantum theory, it is
worthy to study this vital problem. Based on the experimental works
of direct measurement of single two dimensional systems and its density
operators\citep{Thekkadath2016}, and direct measurement of pure two
entangled photon state\citep{Pan2019}, we anticipate that in the
near future the experts can do experiments by taking those two innovative
works into account, and realize the direct measurement of density
operator by considering the theoretical results of our current work.
In our schemes any matrix elements of an entangled state can be obtained
efficiently via proper weak and strong measurements. Since the density
operator representation of a quantum state is general than the wave
function, if consider the open system cases it suggested that the
determination of density operator of an unknown two photon entangled
state may have more practical applications rather than the direct
measurement of complex probability amplitude of state vector of a
pure two photon entangled sate. 
\begin{acknowledgments}
 This work was supported by the National Natural Science Foundation
of China (Grant No. 11865017, No.11864042).
\end{acknowledgments}

\bibliographystyle{apsrev4-1}
\addcontentsline{toc}{section}{\refname}\bibliography{ref}

\end{document}